\documentclass[preprint]{aastex}
\usepackage{amsmath, amsfonts, amssymb,natbib}
\begin{document}

\title{KK246, a dwarf galaxy with extended H \textsc{I} disk in the Local Void}

\author{K. Kreckel\altaffilmark{1}, P. J. E. Peebles\altaffilmark{2}, J. H. van Gorkom\altaffilmark{1}, R. van de Weygaert\altaffilmark{3}, J. M. van der Hulst\altaffilmark{3}}

\altaffiltext{1}{Department of Astronomy, Columbia University, Mail Code 5246, 550 West 120th Street, New York, NY 10027, USA; email: kstanonik@astro.columbia.edu}
\altaffiltext{2}{Joseph Henry Laboratories, Princeton University, Princeton, NJ 08544, USA}
\altaffiltext{3}{Kapteyn Astronomical Institute, University of Groningen, PO Box 800, 9700 AV Groningen, the Netherlands}

\begin{abstract}
We have found that KK 246, the only confirmed galaxy located within the nearby Tully Void, is a dwarf galaxy with an extremely extended H \textsc{i} disk and signs of an H \textsc{i} cloud with anomalous velocity.  It also exhibits clear misalignment between the kinematical major and minor axes, indicative of an oval distortion, and a general misalignment between the H \textsc{i} and optical major axes. We measure a H \textsc{i} mass of $1.05 \pm 0.08 \times 10^8 M_\sun$, and a H \textsc{i} extent 5 times that of the stellar disk, one of the most extended H \textsc{i} disks known.  We estimate a dynamical mass of $4.1 \times 10^9 M_\sun$, making this also one of the darkest galaxies known, with a mass-to-light ratio of 89.  The relative isolation and extreme underdense environment make this an interesting case for examining the role of gas accretion in galaxy evolution.

\end{abstract}

\section{Introduction} \label{intro}

The lowest-density environments, voids \citep{Einasto1980, Kirshner1981}, provide an interesting location to study the gas content and accretion history of galaxies \citep{weyplaten2009}.  Despite their underdense environment, galaxies in voids are bluer with more ongoing star formation than field galaxies \citep{Rojas2004, Rojas2005, Park2007, BendaBeck2008}. In H \textsc{i}, they are typically late type and gas rich \citep{Szomoru1996, Kreckel2011}.  As they appear to reside in a generally more youthful state, diffuse gas accretion may still be a dominant process affecting void galaxies in low mass halos.  \cite{Keres2005} specifically predict cold mode accretion to dominate in low-density environments.

KK 246, also referred to as ESO 461-036, is a dwarf irregular galaxy residing a few Mpc within the Tully Void, an extremely large void at least 23 Mpc in radius \citep{Tully2008}.  It is extremely isolated, with no companion galaxies discovered within 3 Mpc either optically \citep{Karachentsev2004} or in neutral hydrogen (HIPASS, \citealt{Wong2006}).  Its M$_{\rm H~\textsc{i}}$/L$_B$ ratio does not distinguish it from other dwarf irregular galaxies \citep{Karachentsev2006}.  It is the only confirmed galaxy to reside in the Local Void, and one of only 16 galaxies tentatively identified to reside in this enormous volume \citep{Nasonova2010}, presenting a challenge to $\Lambda$CDM, which overpopulates the voids \citep{Peebles2001, Tinker2009, Tikhonov2009}.

 This galaxy was targeted as part of the Faint Irregular Galaxies GMRT Survey (FIGGS, \citealt{Begum2008}), where GMRT H \textsc{i} imaging shows disk-like rotation and the suggestion of a huge warp at the faintest contours.  However, as the galaxy fell at the edge of the observing band, imaging did not include the full velocity range of this galaxy. 
We have imaged KK 246 at the VLA and EVLA with high sensitivity and discovered it to have an extremely extended H \textsc{i} disk as well as possible signs of ongoing gas accretion.  In Section \ref{obs} we describe the observations and  in Section \ref{results} we discuss the results.  In Section \ref{discussion} we discuss the implications and in Section \ref{conclusions} we present our conclusions.

\section{Observations} \label{obs}

As our target is at low declination, it was first observed  at the VLA in CnB array during spring 2009.  These observations were centered at slightly higher frequencies to avoid the effects of aliasing in the lower 0.5 MHz of the bandpass by the VLA correlator on the EVLA antennas, and our target is well contained within the remaining unaffected 2.625 MHz bandwidth.  A correlator problem with antenna 12, which was needed for the shortest baseline, led to extended and irregular sidelobes and poor sensitivity to extended emission.  We supplemented these data in summer 2010 with EVLA observations in D array.    Observing parameters for both array configurations are listed in Table \ref{tab:obsparams}. The D array observations were made at a lower velocity resolution, but the data were resampled to match the velocity distribution of the CnB observations so the two data sets could be combined in the UV plane.  Continuum emission was removed by interpolating from the line free channels in the UV plane. Images were made with natural weighting and a 30 arcsecond UV taper to achieve the highest sensitivity.  After Hanning smoothing we have a velocity resolution of 10.3 km s$^{-1}$; the spatial resolution is 47\arcsec $\times$ 33\arcsec.  Using the Clean algorithm, we constructed an image cube where we cleaned down to 0.4 mJy beam$^{-1}$ (1$\sigma$) using a clean box around the emission in each channel.  Total intensity and intensity weighted velocity field maps were constructed with a smoothing mask of 30 km  s$^{-1}$, 60\arcsec, and a 2.5$\sigma$ cutoff.  The calibration of the VLA data was done in AIPS, and the calibration of the EVLA data was done in CASA.  Both data sets were combined, cleaned and imaged in CASA.

\section{Results} \label{results}

\begin{figure}[h!]
\centering
\includegraphics[height=3in]{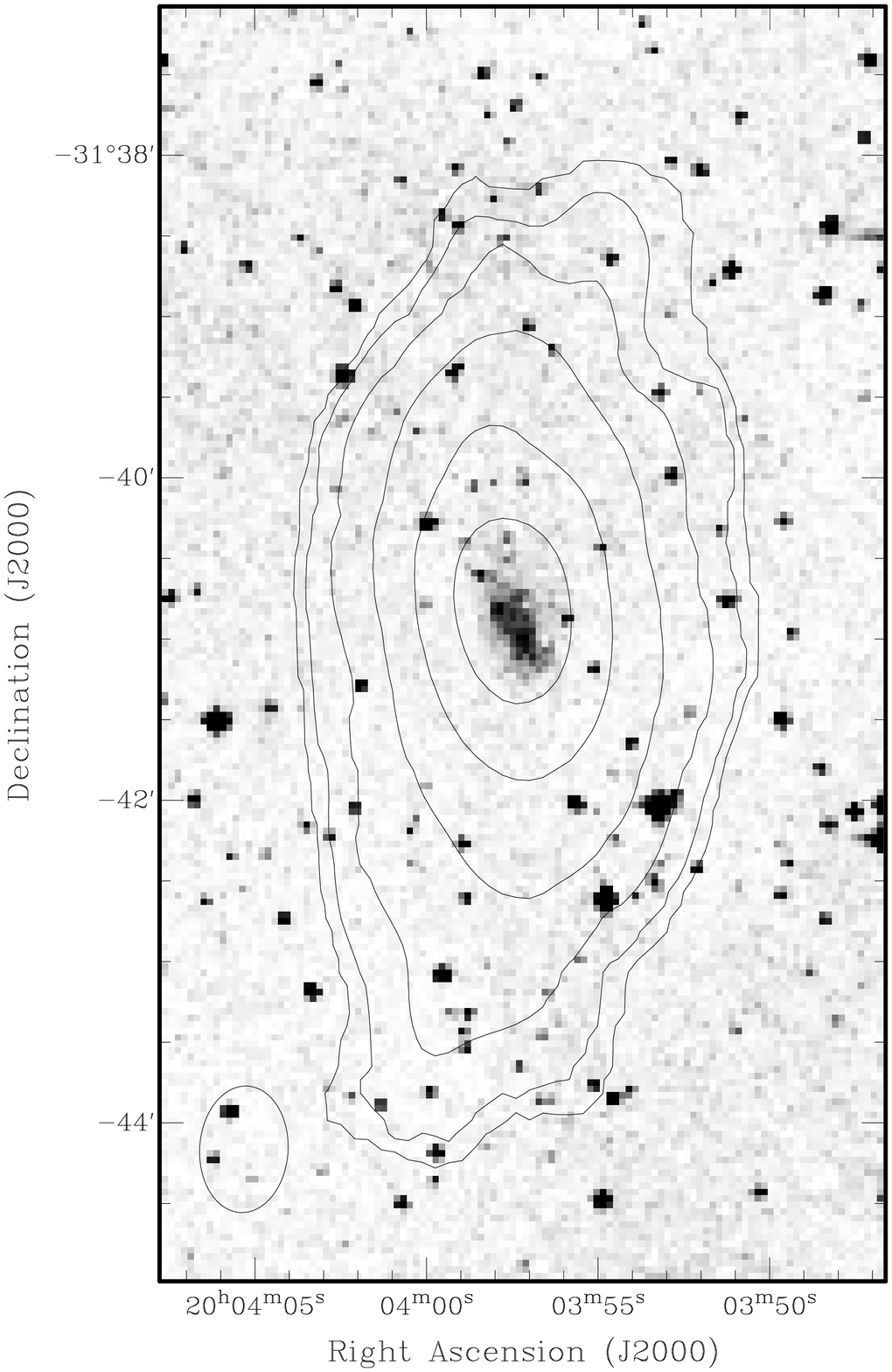}
\includegraphics[height=3in]{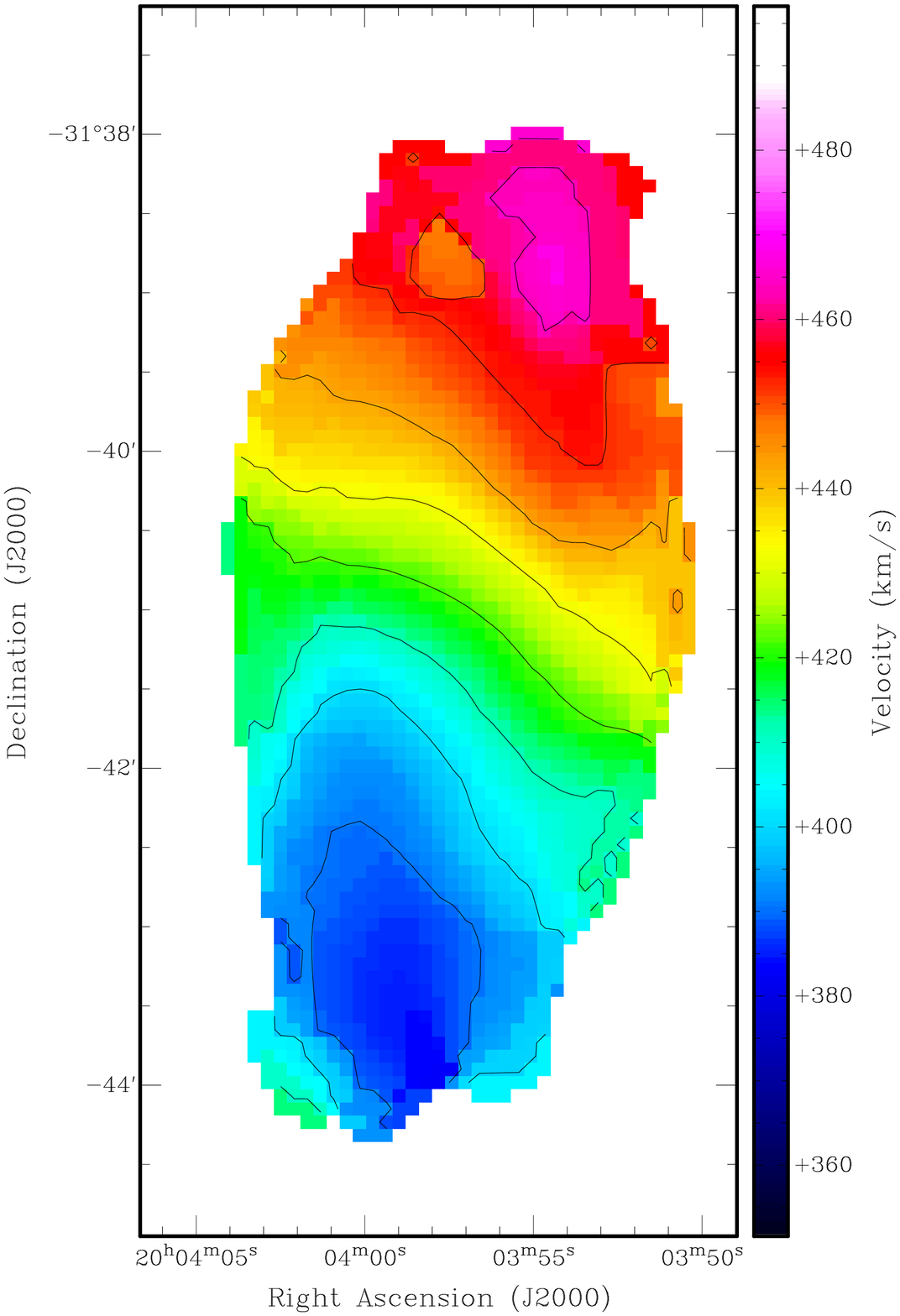}
\raisebox{1.2cm}{\includegraphics[height=2in]{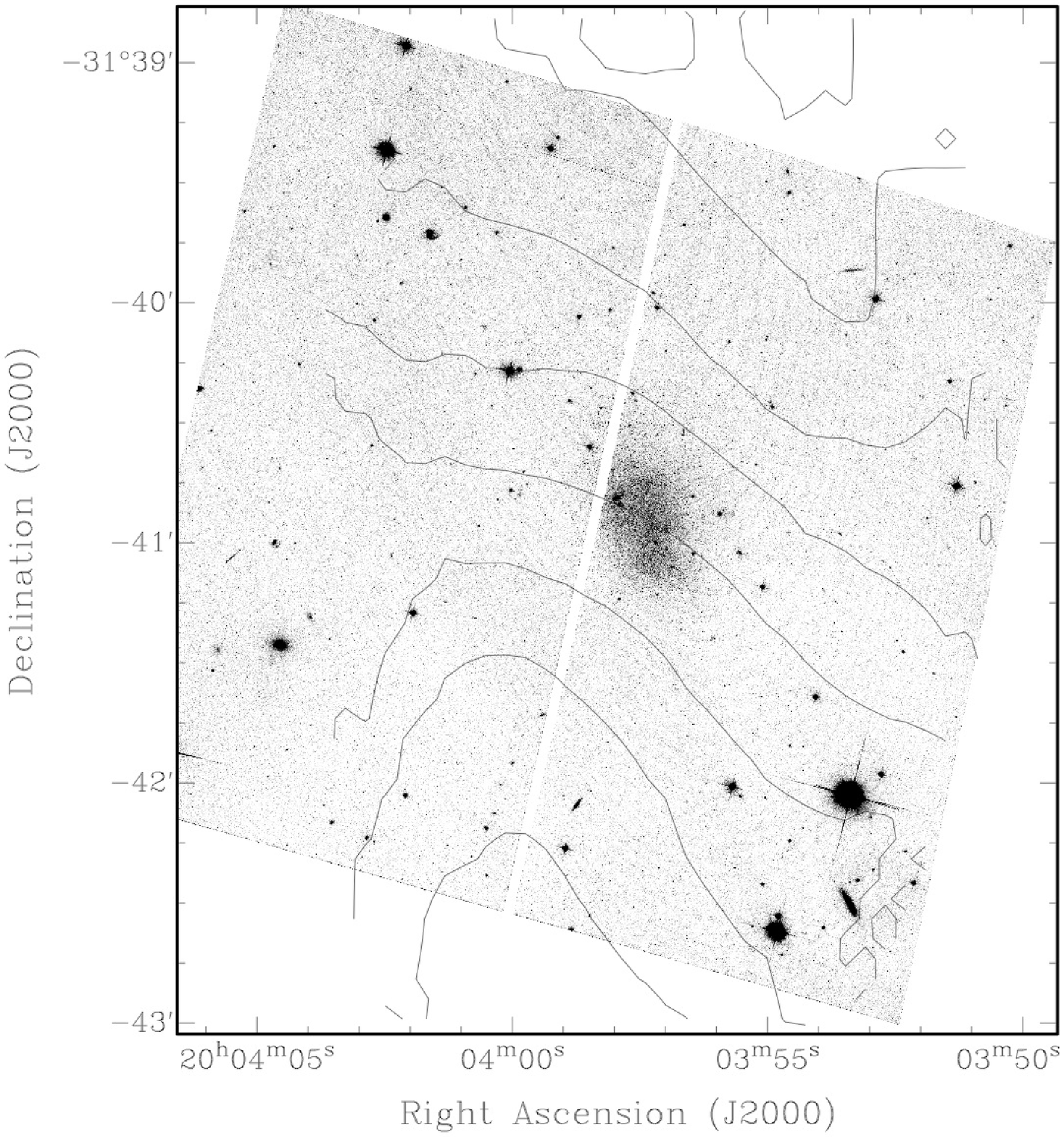}} 
\caption{Left: KK 246, B-band Second Palomar Observatory Sky Survey Digitized Sky Survey image, overlaid with H \textsc{i} contours.  The column density contours are  $2 (1.8\sigma), 4, 8, 16, 32 \times 10^{19}$ cm$^{-2}$.  Center: the velocity field, with increments of 10.3 km  s$^{-1}$ marked.  Right: HST ACS F606W image from the Hubble Legacy Archive overlaid with the velocity field.  
\label{fig:kklo}}
\end{figure}

H \textsc{i} contours overlaid on the Second Palomar Observatory Sky Survey Digitized Sky Survey B-band image, the corresponding velocity field, and the velocity field overlaid on the HST ACS F606W image from the Hubble Legacy Archive are shown in Figure \ref{fig:kklo}.  Most apparent is the misalignment between the major axes of the H \textsc{i} and optical disks. This alignment is not quite a polar disk, exhibiting an angle of about 20$^\circ$ in the center and almost 55$^\circ$ at the largest extent of the H \textsc{i} disk.

We measure a total H \textsc{i} mass of 1.05 $\pm$ 0.08 $\times 10^8 M_\sun$, assuming a distance of 7.83 Mpc \citep{Karachentsev2004}, and a total flux of 7.3 Jy km  s$^{-1}$ (Table \ref{tab:hiparams}).  This is consistent with the HIPASS observations of 7.5 Jy km  s$^{-1}$ \citep{Doyle2005}.  \cite{Karachentsev2004} measured an absolute B-band magnitude of -12.96 assuming a distance of 5.6 Mpc, however with a revised distance measured using the tip of the red giant branch of 7.83 Mpc \citep{Karachentsev2006} this becomes -13.69 magnitudes, and gives a H \textsc{i} mass to light ratio of  M$_{\rm H~\textsc{i}}$/L$_B$ = 2.32.

\begin{figure}[h!]
\centering
\includegraphics[width=4in]{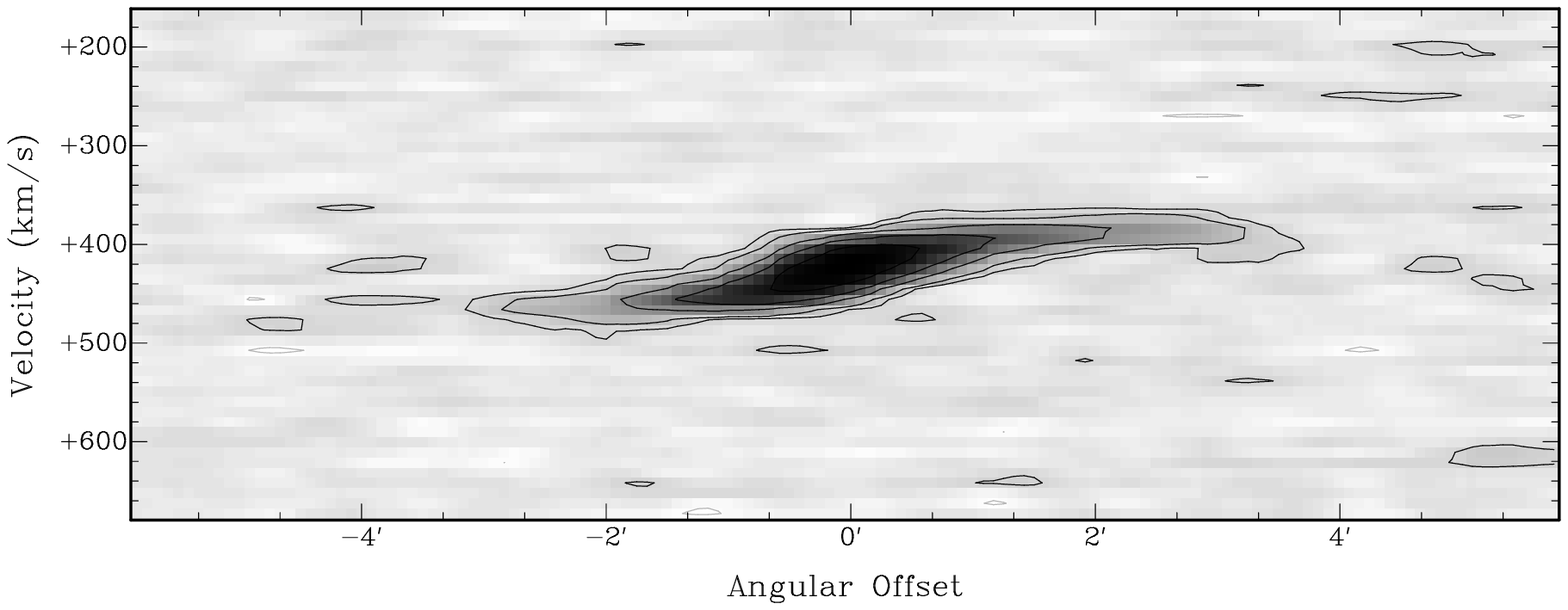}
\includegraphics[width=4in]{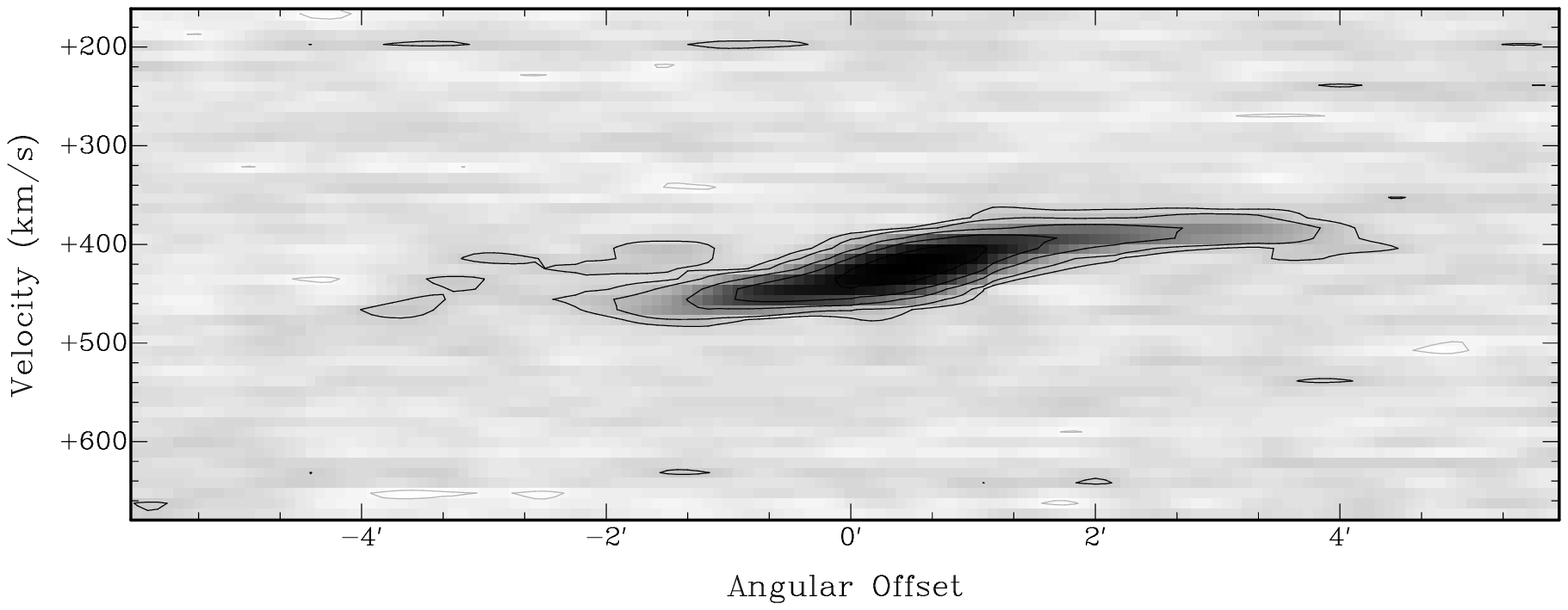}
\caption{Position-velocity diagrams, aligned with the major axis (top) and annomalous gas (bottom).  Contours are at -1, 1 (2.5$\sigma$), 2, 5,  10, and 20 mJy beam$^{-1}$.   
\label{fig:kkpv}}
\end{figure}

\begin{figure}[h!]
\centering
\includegraphics[height=6in]{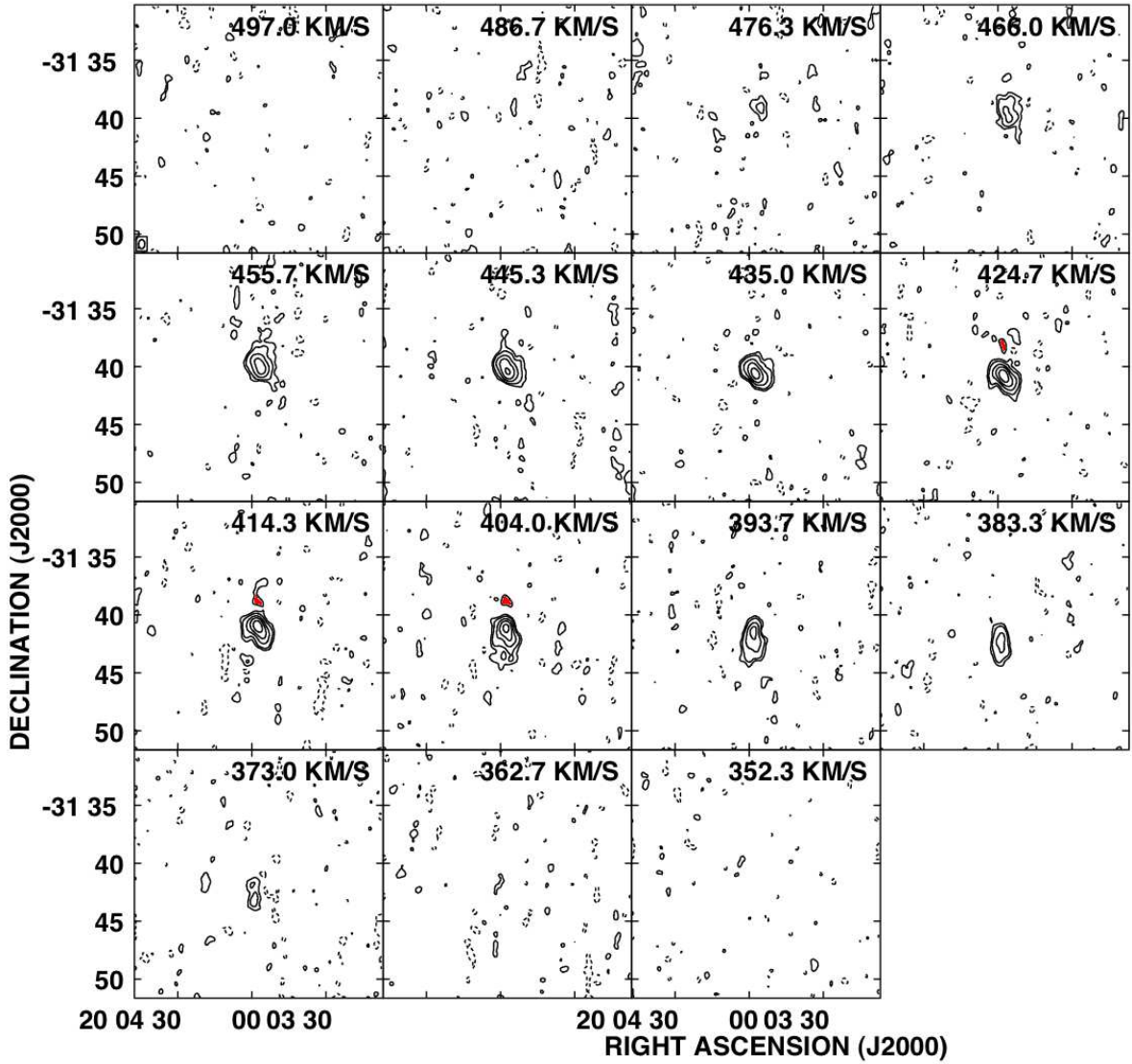}
\caption{H \textsc{i} channel maps.  Contours are -1, 1 (2.5$\sigma$), 2, 5,  10, and 20 mJy beam$^{-1}$.   The synthesized beam is in the lower left corner of the top left image.  The anomalous gas clump can be seen in three channels and is shaded in red.
\label{fig:channels}}
\end{figure}

The H \textsc{i} velocity field exhibits a noticable misalignment of the kinematic major and minor axes (Figure \ref{fig:kklo}, right panel). This signature could be caused by departure from axisymmetry in the mass distribution, such as an oval distortion and 
is indicative of a bar instability that may drive gas inflow \citep{Bosma1981}.  A position velocity slice along the kinematic disk major axis (position angle $-10^\circ$ degrees) shows regular disk rotation with a flat rotation curve (Figure \ref{fig:kkpv}, top panel).   Apparent in the velocity field is a spot on the northeast edge of the disk with an anomalously low velocity compared with regular disk rotation.  A position velocity slice from the disk center through this position (Figure \ref{fig:kkpv}, bottom panel) shows a faint, 2.5 $\sigma$ feature that is extended in both position and velocity.  Careful examination of the H \textsc{i} channel maps (Figure \ref{fig:channels}) reveals that the emission (indicated in red) persists over 20 km  s$^{-1}$ in three separate channels.  Any emission in channels at higher velocities is masked by emission from the bulk of the disk gas.  

\begin{figure}[h!]
\centering
\includegraphics[height=1.8in,angle=0]{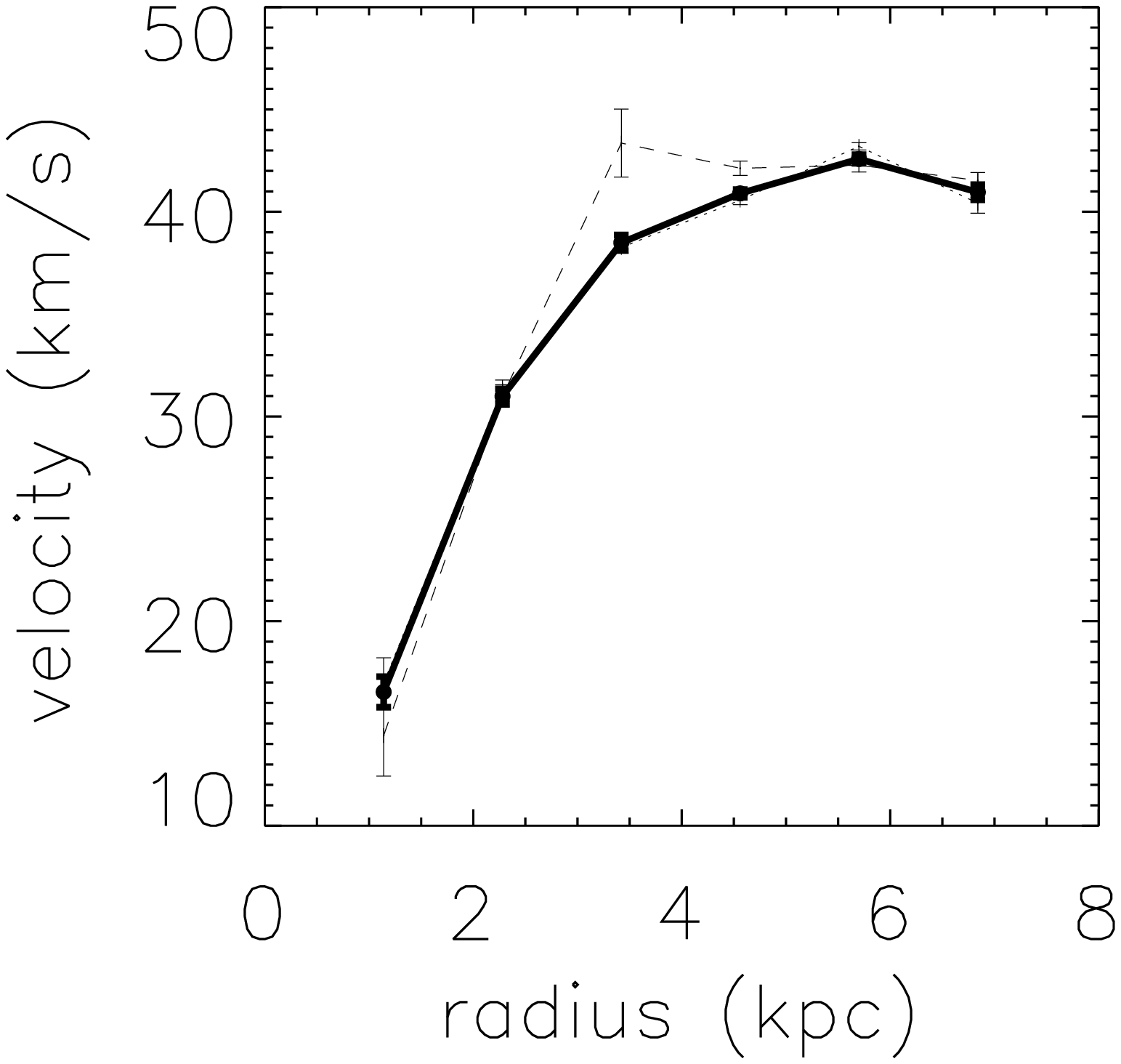}
\includegraphics[height=1.8in,angle=0]{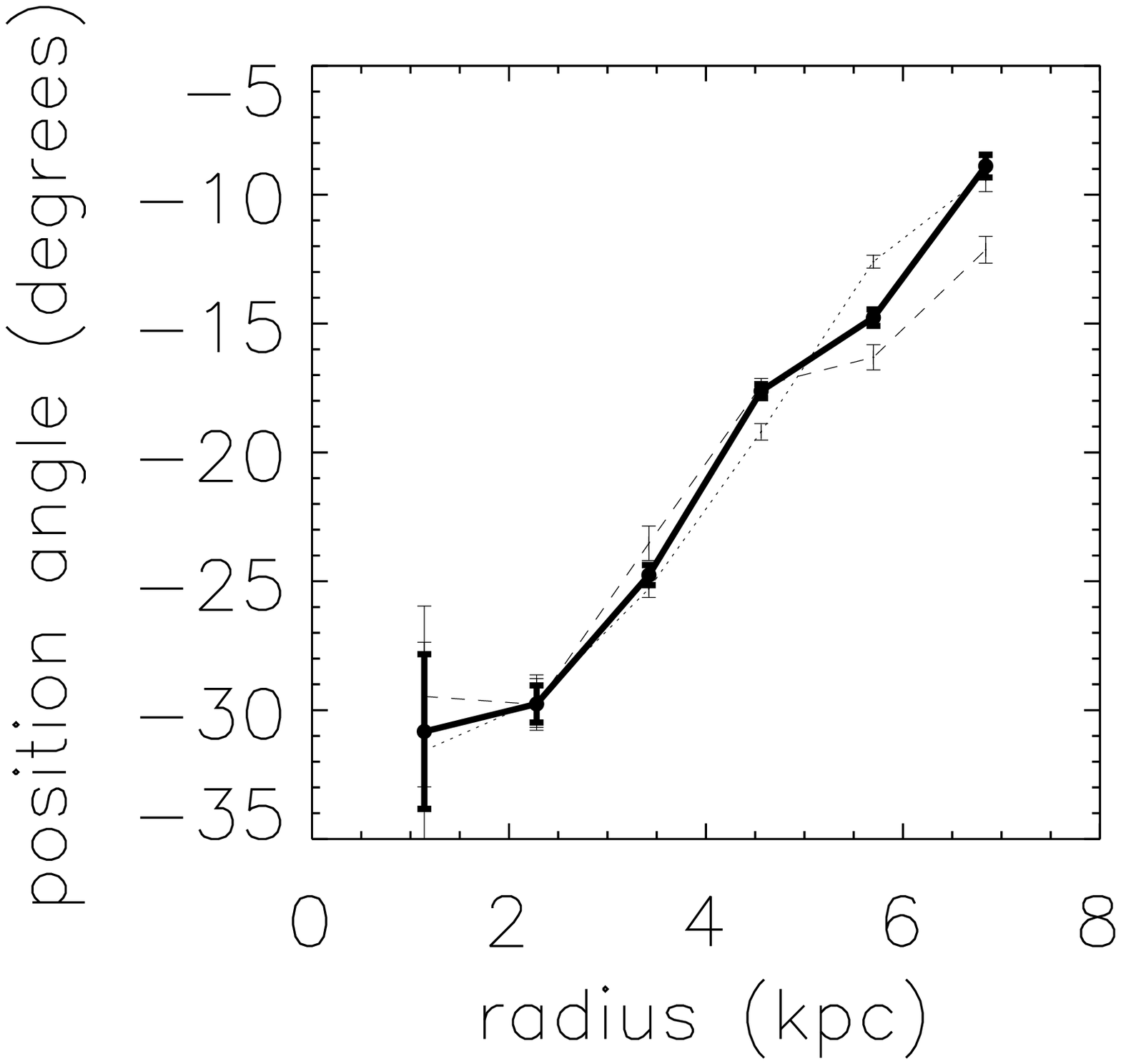}
\includegraphics[height=1.8in,angle=0]{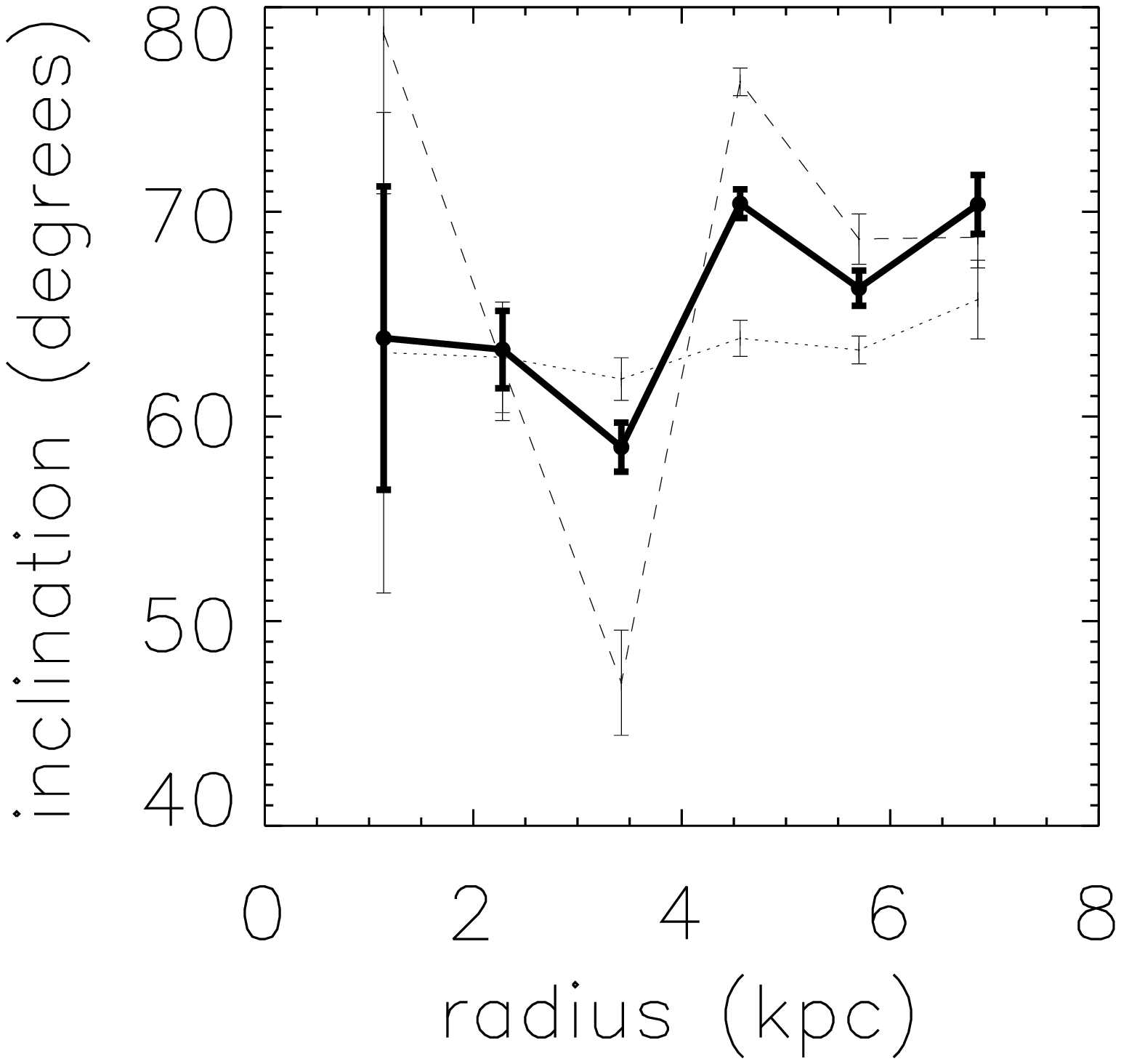}
\caption{Tilted ring fit parameters, calculated using  ROTCUR, for the velocity (left), position angle (center) and inclination (right) as a function of radius.  Values calculated for the whole disk are shown with a solid line.  The approaching side (dotted line) and the receding side (dashed line) calculated separately are also shown.
\label{fig:rotcur}}
\end{figure}

\begin{figure}[h!]
\centering
\includegraphics[height=4in,angle=90]{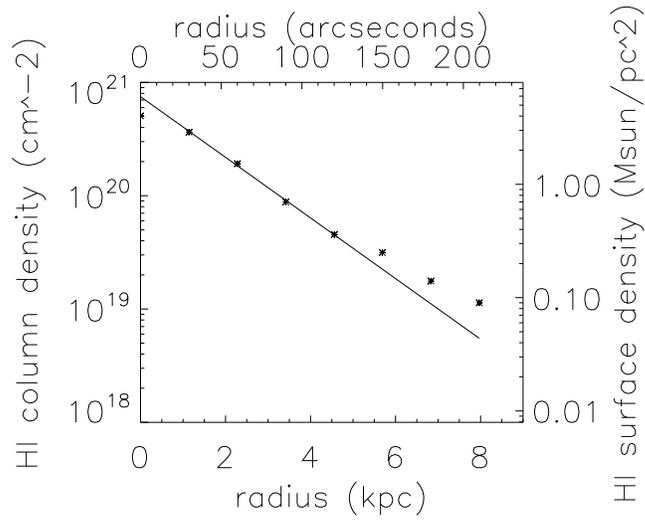}
\caption{Surface density profile using output from ROTCUR in ELLINT.  Values are corrected for face-on, and the last point is fairly low confidence.  The line shows an exponential fit to the inner disk surface density.
\label{fig:surfdens}}
\end{figure}

\begin{figure}[h!]
\centering
\includegraphics[height=3in]{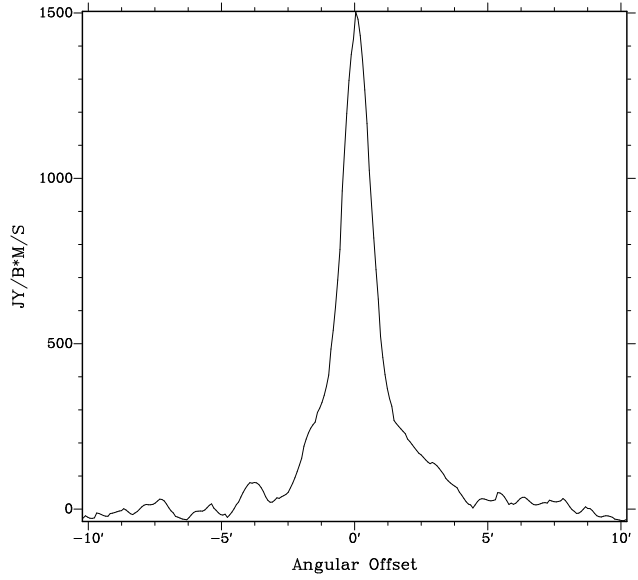}
\caption{Intensity map profile along a position angle of -10$^\circ$, roughly the H \textsc{i} major axis at the furthest extent of the disk.  There is no sharp edge to the disk to the north (right) like we see to the south (left).  Compare with Carignan \& Purton 1998 Figure 4.
\label{fig:wing}}
\end{figure}

The H \textsc{i} disk at its lowest column density contours is quite irregular and somewhat asymmetric.  A tilted ring fit was made to the velocity field using the GIPSY task ROTCUR.  Resulting fit parameters are shown in Figure \ref{fig:rotcur}, however because of the signs of radial streaming motions within the disk the physical implications of the derived position angles and inclinations are unclear.  If we allow for radial motions within the rings, they are best fit by including inward velocities of $\sim$5 -  10 km s$^{-1}$, consistent with streaming motions seen in other disk galaxies \citep{Gentile2007}.
From the global velocity profile, we measure a 20\% H \textsc{i} line width of W$_{20}$ = 93 km  s$^{-1}$.  Correcting for inclination, at an average inclination angle of 65$^\circ$, we find a half-line width W$_{20, i}$/2 = 51 km  s$^{-1}$.  The orientation parameters from the rotation curve analysis were then applied to the total intensity map to calculate the H \textsc{i} surface density profile,  corrected to face-on, which is perfectly exponential in the central disk but somewhat shallower at the furthest extent (Figure \ref{fig:surfdens}).  H \textsc{i} can be reliably detected to a radius of  3\arcmin ~(6.8 kpc at the assumed distance), where the H \textsc{i} face-on column density corresponds to $2 \times 10^{19} ~cm^{-2}$.  Enclosed within this radius we measure a total dynamical mass of $4.1 \times 10^9 ~M_\sun$.  Using the Holmberg diameter D$_{\rm Ho}$ = 1.2\arcmin  \citep{Karachentsev2004} we find a H~\textsc{i} to optical diameter of D$_{\rm H~\textsc{i}}$/D$_{\rm Ho}$ = 5.  Careful inspection of the data suggests that  low column density gas continues to extend to the north, and no sharp edge is seen in a slice in the intensity map profile  (Figure \ref{fig:wing}).  This emission is not at the position of the anomalous gas cloud, but is on the same northern side of the disk. 

\section{Discussion}\label{discussion}

\begin{figure}[h!]
\centering
\includegraphics[height=3in]{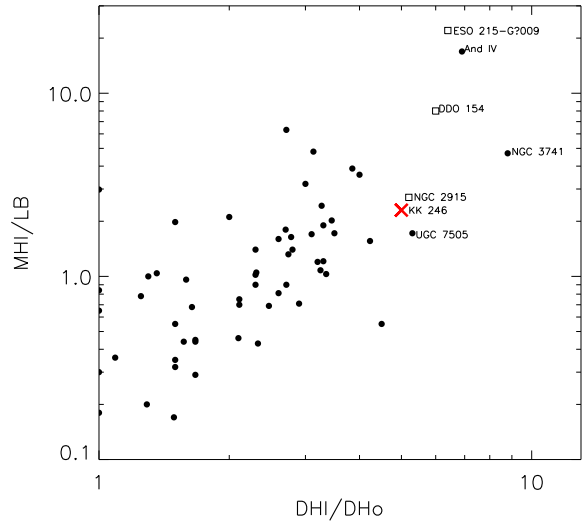}
\caption{H \textsc{i} mass to B-band Luminosity as a function of the H \textsc{i} to optical diameter for the FIGGS sample (dots). KK 246 is marked by a cross, and it and other extended disks from the FIGGS sample are labeled.  Also included are extended H \textsc{i} disks from the literature ESO 215-G?009 \citep{Warren2004}, DDO 154 \citep{Carignan1998}, and NGC 2915 \citep{Meurer1996}.
\label{fig:mhilight}}
\end{figure}

KK 246 is not overly H \textsc{i} rich for its luminosity when compared with the other dwarf galaxies, but its H \textsc{i} disk is quite extended (Figure \ref{fig:mhilight}).  It is similar in extent to other extremely extended H \textsc{i} disk galaxies in the literature (NGC 2915, \citealt{Meurer1996};  DDO 154, \citealt{Carignan1998}; ESO 215-G?009, \citealt{Warren2004}), and the most extended H \textsc{i} disks in the FIGGS sample (And IV; NGC 3741, \citealt{Begum2005}; UGC 7505).  From Figure \ref{fig:mhilight} it is clear that while there is a general trend for more gas rich disks to be more extended compared to their stellar disk, as found by \cite{Begum2008} and \cite{vanZee1995}, it is not necessary that extended H \textsc{i} disks be overly H \textsc{i} rich, or vice-versa.  This suggests that some of these gas disks, like KK 246 and NGC 2915, may be somewhat more diffuse despite their relative size.

\begin{figure}[h!]
\centering
\includegraphics[height=3.5in]{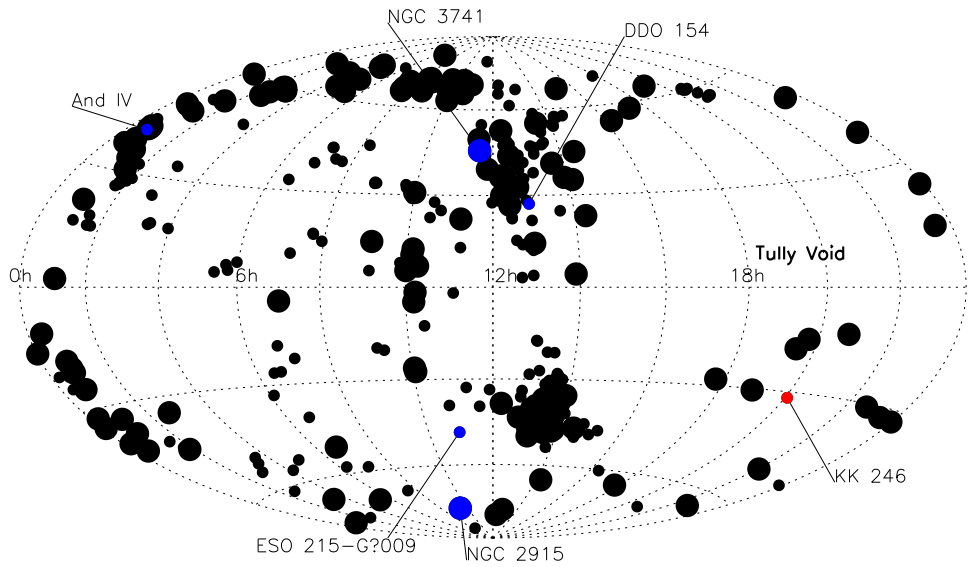}
\caption{All-sky distribution of those galaxies brighter than m$_B$=17.5 within 4 Mpc (larger dots) and within 8 Mpc (smaller dots).  The extended H \textsc{i} disk galaxies discussed in the text, as well as the Local Void, are marked accordingly.
\label{fig:aitoff}}
\end{figure}

Most of these other extended H \textsc{i} galaxies are described as `isolated' by the literature, and for comparison we look in detail at the environments of these galaxies within the Local Volume and around the Local Void (Figure \ref{fig:aitoff}).  \cite{Karachentsev2004} has a catalog of local galaxies that is $\sim$80\% complete to a limiting apparent magnitude of m$_b$ = 17.5, roughly an absolute magnitude of -12 a distance of 8 Mpc, with the majority of the distances determined by Hubble flow-independent measures. We also take into account  improved distances to some of the sample \citep{Karachentsev2006}.  Except for UGC 7505, which is at 12 Mpc, and KK246, which is nearly at 8 Mpc but is well established to be within the Local Void \citep{Tully2008}, all are well within the boundaries of this catalog.  The mean nearest neighbor distance is about 400 kpc, and by this measure all of these galaxies are relatively isolated, none having a neighbor within 500 kpc.  Here we include all neighbors that are brighter than a limiting magnitude of m$_B=17.5$, approximately M$_B =-14.5$ at 8 Mpc.
This is true for all galaxies, except NGC 3741, even when considering the distance to the sixth nearest neighbor.  This suggests that a galaxy must be relatively undisturbed to develop and sustain such an extended H \textsc{i} disk.  However there is no clear correlation with underdensity, as our void galaxy does not have the most extended disk.  The polar disk galaxy recently discovered within a void is also found to have a strikingly extended H~\textsc{i} disk, 
and in that case the lack of stars and substantial mass in the perpendicular component suggest that cold accretion is the most likely source of the polar gas disk \citep{Stanonik2009}.

It is difficult to make direct comparisons between KK 246 and other void galaxies as its proximity makes it uniquely well resolved.  KK 246 is significantly fainter than other void galaxies imaged in H \textsc{i} \citep{Szomoru1996,Kreckel2011}, however it is roughly comparable to the 6 H \textsc{i} detected companion void galaxies in \cite{Kreckel2011}, which range from $-11 > M_r > -16$ and have H \textsc{i} masses of 5 - 30 $\times 10^7$ M$_\sun$.  KK 246 is not particularly blue in color when compared with low luminosity SDSS galaxies, having B - R $\sim$ 1 \citep{Lauberts1989}, but it is significantly redder than the companion void galaxies, which have $B-R \sim 0.6$.  
While these galaxies also do not have particularly extreme H \textsc{i} mass to light ratios, their H \textsc{I} extents and kinematics are largely unresolved and present an interesting population for higher resolution imaging.

Because the disk of KK 246 is so large, we are able to trace the flat rotation curve out to fairly large radii, and observe no decline at the outmost extent.  Thus we measure a fairly high lower-limit for the total dynamical mass, and relatively large mass-to-light ratio of 89, making this one of the darker galaxies known (other examples are NGC 2915 and NGC 3741 with mass-to-light ratios of 76 and 107 respectively).  An H-band study of local galaxies \citep{Kirby2008} estimated a stellar mass of $5 \times 10^7 M_\odot$, approximately half the H \textsc{i} mass we report.  Assuming a total gas mass of 1.4 times the H \textsc{i} mass, we find an upper limit on the baryon fraction of 5\%, substantially lower than the expected cosmic baryon fraction of 17\%, but in agreement with the finding that the lowest mass halos are `missing' relatively more of their baryons \citep{Mcgaugh2010}.  
Simulations of dwarf galaxies in voids and filaments find strong suppression of star formation through photoheating by the UV background as a function of halo mass and independent of the galaxy environment \citep{Hoeft2006, Hoeft2010}, though for the inclination-corrected half-line width W$_{20,i}$/2 = 51 km  s$^{-1}$ we find that KK 246 is still relatively baryon poor compared to these predictions although  in good agreement with the baryonic Tully-Fisher relation as extended to low-mass dwarf galaxies by \cite{Trachternach2009}. \cite{Hoeft2010} in particular suggest that galaxies at the outskirts of voids, as KK 246 is, may be subject to additional photo-heating from nearby higher density regions and thus be relatively darker than galaxies in the void centers.

It is possible we are missing a diffuse, extended stellar component in the galaxy. High resolution ACS imaging shows KK 246 to be optically quite irregular, with no clear nucleus or disk structure (Figure \ref{fig:kklo}, right panel).  Perhaps the main stellar component we see is a bar for an extremely  low surface brightness stellar disk. This would explain some of the anomalous disk kinematics we see, particularly the misalignment between the major and minor kinematic axes, which can be driven by a bar instability and is typically a sign of gas inflow or outflow.  However, the H \textsc{i} is inclined to nearly edge-on and we see no diffuse stellar features along the H \textsc{i} disk (Figure \ref{fig:kklo}, left), and the galaxy is not particularly under-luminous for the H \textsc{i} we observe (Figure \ref{fig:mhilight}). This  suggests that the kinematic misalignment may be a result of gas inflow without, necessarily, a driving bar.  
With such a small, dark matter dominated galaxy it is possible that the rotation of a triaxial dark matter halo could drive the irregular kinematics and large extent, similar to what is seen in NGC 2915 \citep{Bureau1999}, and sustain a long-lived sloshing of the neutral hydrogen disk that prevents settling to form stars in the disk outskirts.

In general, there is an apparent misalignment between the major axis of the H \textsc{i} and stellar disks, so perhaps the irregular gas disk kinematics is indicative of a large scale warp.  \cite{Roskar2010}  simulated a persistent extended warp in a disk galaxy, and found that a relatively minor merger was able to misalign the stellar disk from the extended hot gas halo, but any subsequently accreted cold gas was torqued into alignment with the original orientation, along the hot halo.  They find their warps are destroyed when the influx of fresh gas is cut off, or when the central stellar disk grows to occupy the region of the warp.  The void environment may be ideal for retarding both these mechanisms, thus allowing the persistence of large extended warped disks.

The anomalous H \textsc{i} cloud we observe is similar to, but less significant than, the H \textsc{i} `beard' observed in NGC 2403, which is believed to result from H \textsc{i} gas located above the plane of the galaxy that is rotating more slowly \citep{Fraternali2002}. 
In KK 246, the anomalous gas is at `forbidden' velocities, inconsistent with the full disk rotation.  
We estimate an upper limit on the H~\textsc{i} mass of $2 \times 10^6 M_\sun$. This is consistent with the mass estimate for high velocity clouds around the Milky Way Galaxy \citep{vanwoerden1999}, and presents a similar situation to the high velocity clouds found around NGC 2403 and M33 that provide the best observational evidence of ongoing gas accretion onto galaxies \citep{Sancisi2008}.  As the cloud in KK 246 is clearly far removed from the small stellar component at the center of the galaxy, it seems that its origin from galactic fountain effects is unlikely. 
In addition, the inner disk is perfectly exponential, but the decrease in surface density at the edge is not as sharp (Figure \ref{fig:surfdens}), which combined with the trailing faint H \textsc{i} emission to the north of the galaxy (Figure \ref{fig:wing}) is suggestive of additional ongoing gas accretion.

\section{Conclusions}\label{conclusions}
We have observed in detail the H \textsc{i} disk of KK 246, the only confirmed galaxy in the Local Void, and found that it is extremely extended and shows signs of an anomalous gas cloud within the disk.  In addition, there is a clear misalignment between the major axis of the H \textsc{i} and stellar disks, and there are kinematic signs of an oval distortion in the H \textsc{i} disk, suggestive of gas inflow in the central region.  We measure a total H \textsc{i} mass of $1.05 \pm 0.08 \times 10^8 M_\sun$, and a H \textsc{i} extent of 5 times the optical Holmberg diameter, one of the most extended H \textsc{i} disks known.  The rotation curve is flat to the edge of the H \textsc{i} disk, and we calculate a dynamical mass of $4.1 \times 10^9 M_\sun$.  Thus it is also one of the darkest galaxies known, with a mass-to-light ratio of 89.

When compared with other galaxies with extended H \textsc{i} disks, we find that relative isolation is a necessary, but not sufficient, condition for their existence.  Alternatively, we find that these extended H \textsc{i} disks are not necessarily overly H \textsc{i} rich for their luminosity, as in the case of KK 246.  This suggests that these systems are able to convert their gas into stars in a fairly typical manner, while retaining an extended diffuse gas disk.  The anomalous kinematics of a small, $< 10^6 M_\sun$, component of the disk suggest that perhaps ongoing gas accretion is playing a role in maintaining this extended disk.

\acknowledgments

KK thanks Tony Wong for useful discussions and valuable comments. We also thank the referee for their helpful comments.  This work was supported in part by the National Science Foundation under grant \#1009476 to Columbia University. The National Radio Astronomy Observatory is a facility of the National Science Foundation operated under cooperative agreement by Associated Universities, Inc.  

Based on observations made with the NASA/ESA Hubble Space Telescope, and obtained from the Hubble Legacy Archive, which is a collaboration between the Space Telescope Science Institute (STScI/NASA), the Space Telescope European Coordinating Facility (ST-ECF/ESA) and the Canadian Astronomy Data Centre (CADC/NRC/CSA).

The Digitized Sky Surveys were produced at the Space Telescope Science Institute under U.S. Government grant NAG W-2166. The images of these surveys are based on photographic data obtained using the Oschin Schmidt Telescope on Palomar Mountain and the UK Schmidt Telescope. The plates were processed into the present compressed digital form with the permission of these institutions.
The Second Palomar Observatory Sky Survey (POSS-II) was made by the California Institute of Technology with funds from the National Science Foundation, the National Geographic Society, the Sloan Foundation, the Samuel Oschin Foundation, and the Eastman Kodak Corporation.

\clearpage

\begin{table}
\begin{center}
\tabletypesize{\scriptsize}
\caption{Parameters of the VLA, EVLA and combined observations
\label{tab:obsparams}}
\begin{tabular}{l l l l}
\tableline
\tableline
\textbf{Quantity}	& 	\textbf{VLA} 		& \textbf{EVLA}  & \textbf{Combined}\\
Configuration 		& 	CnB 		& D & ...\\
Date 			& 	May/June 2009 	& July 2010  & ...\\
No. telescopes 		& 	 25 		& 19  & ...\\
Exposure time (h) 	& 	 18 		& 2.5 &  ...\\
Total bandwidth (MHz) & 	 3.1  & 4  & 3.1 \\
Channel width (kHz)		& 	24.4 & 15.6  & 24.4 \\
No. channels 		& 	128 		& 256 & 128 \\
Shortest spacing (m)	& 	35		&  35 & 35 \\
Longest spacing (km)	& 	$\sim$10		& 1 & 10 \\
FWHP primary beam (arcmin) & 	$\sim$32   & $\sim$32  & $\sim$32  \\
Synthesized beam (arcsec $\times$ arcsec)&  16 $\times$ 14 & 131  $\times$ 43  & 47  $\times$ 33 \\
Noise per channel (mJy beam$^{-1}$) & 0.5  & 1.7   & 0.4  \\
\tableline
\end{tabular}
\end{center}
\end{table}

\begin{table}
\begin{center}
\tabletypesize{\scriptsize}
\caption{Optical and H \textsc{i} Parameters
\label{tab:hiparams}}
\begin{tabular}{l l l}
\tableline
\tableline
 \textbf{Parameter}		& 	\textbf{Value}  		& \textbf{Reference}\\
R.A. (J2000)		&	20 03 57.4	& \cite{Lauberts1982}\\
Dec (J2000)		&	-31 40 53		& \cite{Lauberts1982}\\
Dist (Mpc)			& 	7.83 	& \cite{Karachentsev2006} \\
R$_{Ho}$ (arcmin)	&	0.6			& \cite{Karachentsev2004}\\
m$_B$ (mag) 		& 17.06			& \cite{Lauberts1989}\\
B-band extinction (mag)	& 1.28			& \cite{Karachentsev2004} \\
M$_B$ (mag)		& -13.69		& ... \\
L$_B$ ( L$_\sun$)	& 	$4.6 \times 10^7$ & ... \\
Total H \textsc{i} flux (Jy km  s$^{-1}$)		&	7.3	& this work\\
Total H \textsc{i} mass ( M$_\sun$)		&	$1.05 \pm 0.08 \times 10^8$& this work\\
R$_{\rm H~\textsc{i}}$ (arcmin)			&	3 			& this work\\ 
W$_{20}$ (km  s$^{-1}$)			&	93 		& this work\\
M$_{dyn}$ ( M$_\sun$)		&	$4.1 \times 10^9$ & this work\\
M$_{\rm H~\textsc{i}}$/L$_B$	&	2.3			& this work\\
R$_{\rm H~\textsc{i}}$/R$_{Ho}$	&	5			& this work\\
M$_{dyn}$/L$_B$	&	89			& this work\\
\tableline
\end{tabular}
\end{center}
\end{table}

\clearpage

\end{document}